\documentclass[aps,pre,twocolumn]{revtex4-1}
\usepackage{graphicx,esvect,mathtools,physics,lineno,bm}
\usepackage{hyperref}
\usepackage{xcolor}
\usepackage{amsmath}
\usepackage{amsfonts}
\usepackage{amssymb}
\usepackage{varioref}
\usepackage{float}
\usepackage{dcolumn}   %needed for some tables
\usepackage{subfigure}

\begin{document}
\title{Aggregate morphology of active Brownian particles on porous, circular walls}
\author{Suchismita Das$^\dagger$, Sounok Ghosh$^\dagger$, Raghunath Chelakkot}
\email{raghu@phy.iitb.ac.in\\
$^\dagger$ Authors contributed equally to this work.}
\affiliation{
 Department of Physics, Indian Institute of Technology Bombay, Mumbai, India.\\ 
}
\begin{abstract}
We study the motility-induced aggregation of active Brownian particles (ABPs) on a porous, circular wall. We observe that the morphology of aggregated dense-phase on a static wall depends on the wall porosity, particle motility, and the radius of the circular wall. Our analysis reveals two morphologically distinct, dense aggregates; a connected dense cluster that spreads uniformly on the circular wall, and a localized cluster that breaks the rotational symmetry of the system. These distinct morphological states are similar to the macroscopic structures observed in aggregates on planar, porous walls. We systematically analyze the parameter regimes where the different morphological states are observed. We further extend our analysis to motile circular rings. We show that the motile ring propels almost ballistically due to the force applied by the active particles when they form a localized cluster, whereas it moves diffusively when the active particles form a continuous cluster. This property demonstrates the possibility of extracting useful work from a system of ABPs, even without artificially breaking the rotational symmetry. 
\end{abstract}

\maketitle

\section{Introduction}

Active fluids that consist of self-motile elements are known for their large-scale collective ordering. A large class of such systems that occur in nature, such as bird flocks, fish schools, and cytoskeletal filaments, display both orientational and density ordering~\cite{Marchetti2013, Ramaswamy2010, Redner2013, Fily2012, Klamser2018, Cugliandolo2017, Farage2015, Cates2015, herminghaus2017phase, vanDamme2019}. However, synthetic self-motile particles without shape anisotropy or alignment interactions show only density ordering. Such a density ordering caused by the aggregation of particles in the absence of adhesive interactions is known as motility-induced phase separation (MIPS) and have been extensively studied theoretically~\cite{Redner2013, Fily2012, Klamser2018, Cugliandolo2017, Farage2015, Cates2015, herminghaus2017phase, vanDamme2019, caprini2020spontaneous},
as well as validated in experiments \cite{Therkauff2012, Jiang2010, Buttinoni2013, Palacci2013, vanderLinden2019, Volpe2011, Bricard2013, Nishiguchi2015, Yan2016, Liu2019, Geyer2019}. Although MIPS is a highly nonequilibrium aggregation process, its qualitative similarities to equilibrium liquid-gas phase separation have motivated the formulation of an effective thermodynamic approach for explaining this phenomenon. Such descriptions involve formulations of pressure\cite{yang2014aggregation, Takatori2014, Solon2015, Winkler2015, Fily_2017, Solon_2018, Speck2016, Patch2017, Nikola2016, Marconi2016}, surface tension\cite{Bialke2015, paliwal2017non, prymidis2016vapour, Junco_JCP2019_surfacetension}, and chemical potential\cite{paliwal2018chemical, takatori2015towards, guioth_jcp2019}, in the context of active systems.

It has been shown previously that the presence of a nonadhesive planar wall enhances the MIPS since the slowing down of active particles due to wall repulsion leads to the nucleation of dense phase on the wall~\cite{Lee_SoftMatter_2017, yang2014aggregation, sepulveda2017wetting, SepulvedaSoto_PRE2018_activewetting}. Since most of the micron-scaled active systems in nature are confined in space, it becomes imperative to understand how the dynamics of these active particles get modified in the presence of such walls\cite{bechinger2016active, Lauga2006, das2018confined}. It has been shown recently that the wall-adhesion properties of active particles can be utilized for sorting and trapping of such particles, by effectively tuning the wall geometry
~\cite{Kaiser2012, Kumar2019, Magiera_PRE2015_trapping, Wu_2018_transport_barrier, Reichhardt_PRE2018_laning}. Besides, changing the wall penetrability of particles also changes their properties near the wall, which has potential applications in specific biomedical processes involving drug delivery~\cite{Daddi_Moussa_Ider_2019, daddi2019membrane}. In the case of nonplanar walls, it has been shown that the wall curvature affects the aggregation properties of active particles~\cite{Nikola2016,smallenburg2015swim,wysocki2015giant}.

%Stochastic thermodynamics\cite{Seifert_2012, Sekimoto_1998} has made it possible to define work at a regime in which fluctuation dominates. There have been experimental\cite{Krishnamurthy2016, Blickle2012, Martinez2016} studies on microscopic heat engines as well as theoretical studies concerning active baths\cite{Zakine_2017, Martin_2018}\cite{Wang_PRL_2002}. 
\begin{figure}
\includegraphics[width=0.45\textwidth]{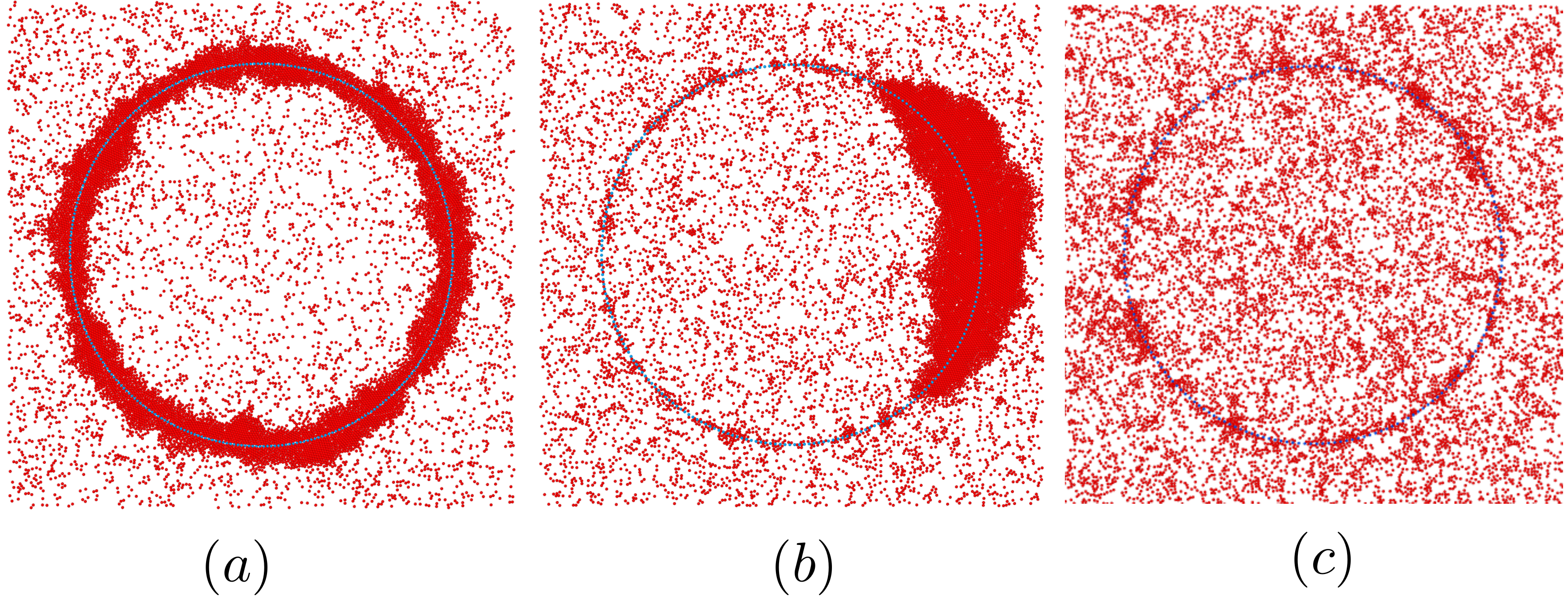}
\caption{Observed variation in clustering behavior at a fixed radius of the circular wall $R = 64$, and $Pe=180$ for different values of pore sizes $\delta$. Increasing values of $\delta$ cause the clustering on the ring to change from (a) a connected dense phase at $\delta = 1.25$ (ESM Movie-1) to (b) a localized cluster dominated by a single large cluster at $\delta = 1.65$ (ESM Movie-2) and finally (c) rapidly fluctuating small clusters at $\delta = 2.05$ (ESM Movie-3). The active particles are colored in red, whereas the static obstacles are colored in blue. All three simulations are performed for a particle area fraction $\phi =0.3$. }
\label{fig:1}
\end{figure}
Because of the nonequilibrium nature of wall adhesion, the active particles continuously propel, thereby exerting mechanical stress on the wall. Similarly, a passive tracer particle immersed in a bath of active particles experiences random drive due to the background activity~\cite{Wu_PRL2000,chen2007fluctuations,leptos2009dynamics,wilson2011differential,mino2011enhanced}. Also, in the presence of an active bath, passive polymers display unusual dynamical behavior~\cite{kaiser2014unusual,harder2014activity,shin2015facilitation,shin2017elasticity,eisenstecken2017internal}. These properties have recently inspired the formulation of active work~\cite{Cagnetta_PRL_2017,  Nemoto_PRE_2019}, and proposed ways of extracting maximum work~\cite{Nemoto_PRE_2019,angelani2009self,sokolov2010swimming}.  Recently, there is a growing interest in studies on the active heat engines both theoretically~\cite{Zakine_2017, Martin_2018, Chaki_PhysicaA_2018, Goswami_2019} %{\color{red} Last two citations related to work fluctuation in the context of active bath} 
and experimentally~\cite{Krishnamurthy2016}. It has been shown that curved~\cite{mallory2014curvature, bechinger2016active}, and chevron shaped~\cite{kaiser2013capturing} passive tracers which break the rotational symmetry can propel ballistically in an active bath, because of the asymmetry in active particle accumulation. 
Recent theoretical investigations are interested in formulating the design principles to optimize the shapes of such passive tracers~\cite{pietzonka2019autonomous}. In our recent study, we have shown the formation of localized clusters of active particles on planar, porous walls~\cite{Das2020}. The same property can be utilized to induce localized clusters on circular walls. The absence of rotational symmetry in such clusters allows us to extract useful work without introducing shape anisotropy. 
 
In this article, we numerically study the aggregation of active particles on a nonadhesive 2D circular wall fixed in space. To make the wall porous, we arrange 'wall' particles at regular intervals, to form a circular ring. 
 We find that the morphology of particle aggregates crucially depends on both the wall porosity and particle motility. Two morphologically distinct clusters are formed as the porosity is increased. One type of cluster spreads entirely along the wall, called connected dense phase (CDP)(Fig \ref{fig:1}(a)), and the other one which does not spread along the entire circumference of the wall is called localized cluster (LC) (Fig \ref{fig:1}(b)), which does not have the rotational symmetry of the adhering wall. When the pore size is much larger than the particle diameter, the aggregation becomes insignificant (Fig \ref{fig:1}(c)). The cluster formation also depends on the radius of the porous circle, especially at moderate porosity, where the cluster disappears below a threshold radius.   
We finally demonstrate that this angular asymmetry in the case of a localized cluster gives rise to force imbalance on a porous circle. When the circle is made motile, this leads to a propulsive motion of the passive, porous circle. Thus, we propose a novel way to extract work from the system without any externally induced anisotropy. We organize the paper as follows: we introduce the model of our system in section II. In section III, we present our simulation results and quantitative analysis for the static circular ring, followed by the nonstatic porous ring in section IV.  We summarize the results in section V.

\section{Numerical model}

We consider a system containing $N$ disk-shaped active particles in 2D, which are enclosed in a square simulation box of length $L$ with periodic boundaries. The particles interact via a repulsive and short-ranged WCA potential, $U = 4\epsilon [(\sigma/r_{ij})^{12} - (\sigma/r_{ij})^6]$ where $r_{ij} < 2^{1/6}\sigma$ and zero otherwise~\cite{WCA1971}, $r_{ij}$ being the distance between particles $i$ and $j$. To model a porous circular wall, we arrange $N_{\text{w}}$ 'wall' particles on a circle of radius $R$ at a regular interval of $d$ in such a way that $N_{\text{w}}d \simeq 2\pi R$. Both $R$ and $d$ are invariant throughout a single simulation.
The active particle interaction potential with the wall $U^{\text{w}}$ is also short-ranged, and repulsive in nature. For the sake of simplicity, we assume $U^{\text{w}} = U$. The time evolution of $i^{th}$ particle position $\textbf{r}_i$ is determined by the overdamped equation,
\begin{equation}
{\dot{\bf{r}}}_i  = D\beta\left(\textbf{F}_i + \textbf{F}^{\text{w}}_i \right) + v_0{\hat{\bf{e}}}_{i}+ \sqrt{2D}{\bm{\eta}}_i, 
\label{eqn:1}
\end{equation}
where $\textbf{F}_i = -\bm{\nabla}_i U$ and $\textbf{F}_i^{\text{w}} = -\bm{\nabla}_i U^{\text{w}}$. $D$ is the diffusion coefficient, $\beta = {1 / k_BT}$, and $\eta$ is Gaussian white noise such that $\langle \eta(t) \rangle$ = 0 and $\langle \eta_{i\alpha}(t)\eta_{j\beta}(t^\prime)=\delta_{ij}\delta_{\alpha\beta}\delta(t-t^\prime)$. $v_0$ denotes the self-propulsive speed of active particles, and $\hat{\bf e}_i = ( \cos\theta_i, \sin\theta_i)$ is the polarity vector, where $\theta_i$  evolves as $\dot{\theta}_{i} = \sqrt{2D_{r}}\eta_{i}^{R}$, $D_r = 3D/\sigma^2$, is the rotational diffusion coefficient. We choose $\sigma$ as the unit for distances, $\tau = {\sigma^2}/{D}$ as the units of time, and $k_BT$ as the unit of energy. 
We use the dimensionless P{\`e}clet number $Pe = {v_0 \sigma}/{D}$ to parametrize the activity, and $\delta = d/\sigma$ is the dimensionless parameter for porosity. Previous studies have shown that the probability of MIPS in the absence of a wall is determined by the ABP motility via $Pe$ and the particle area fraction, $\phi = N a/L^2$, where $a = \pi \sigma^2/4$, the area of a single ABP~\cite{Redner2013, Fily2012}. To focus on the dense-phase formation only on the walls, we choose $\phi = 0.3$ for all our simulations, where we do not observe MIPS in the bulk fluid.  

We study two different systems, one in which the circular wall is static and the second in which the center-of-mass of the wall is allowed to translate (motile walls). In the case of motile walls, we assume the circle to be able to translate and rotate like a rigid body, in response to the active force and the torque applied by the ABPs. Thus, we evolve the center-of-mass of the ring ${\bf \dot{r}}_{\text{cm}} = \mu\sum_{j}^{N_{\text{w}}}\sum_{i}^{N} {\bf -F}^\text{w}_{i,j}(t)$ where the vector summation is over the number of wall particles and $\mu = D\beta/N_{\text{w}}$, the mobility of the circle, and ${\bf F}^{\text w}_{i,j}$ is the interaction force between $i$th ABP and $j$th wall particle. Also, we calculate the net torque ${\bm{\tau}}(t) = \sum_{j}^{N_{\text w}} [{\bf r}_{j}\times\sum_{i}^{N}{\bf -F}^\text{w}_{i,j}(t)]$ upon the ring, where ${\bf r}_{j}$ is the radial vector of $j$th wall particle from the center of the circle. The orientation of the circle is defined by the direction of a body-fixed unit vector, which is pointing to one of the specified wall-particles from the center, denoted by the angle, $\Theta$. The orientation $\Theta$ evolves according to the relation, $\dot{\Theta} = \gamma \tau$, where $\tau$ is the only nonvanishing component of the torque vector, normal to the plane of the circle and $\gamma = \mu/ R^2$.
\begin{figure}
	\includegraphics[width=0.9\columnwidth]{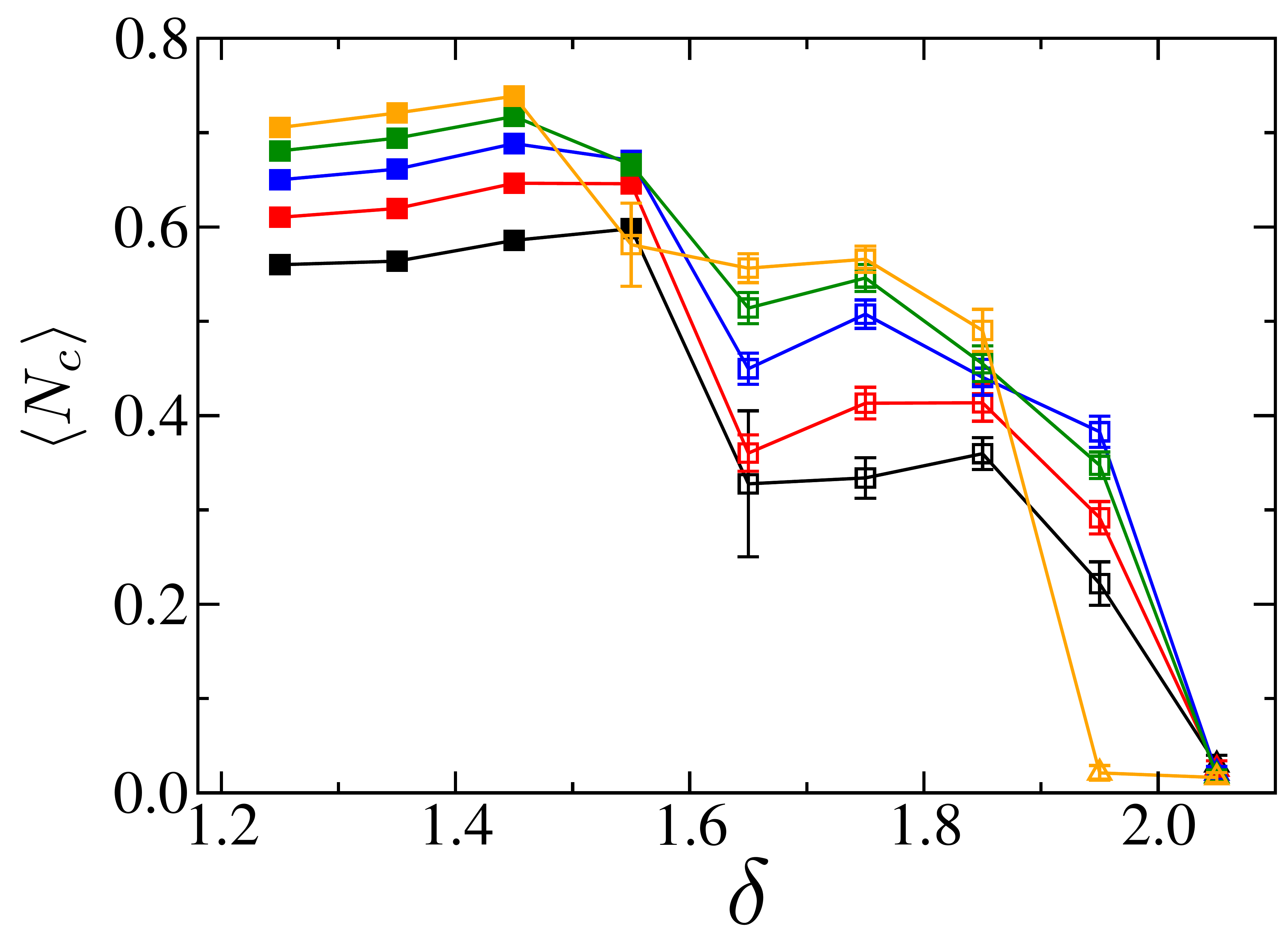}
\caption{The average fraction of ABPs in the dense-phase $\langle N_c \rangle$, as a function of the porosity $\delta$ at $R=64$. Different curves represent different $Pe$. From the bottom on the left, $Pe = 120$ (black), $Pe=150$ (red), $Pe=180$ (blue), $Pe=210$ (green), and $Pe=240$ (orange). The symbol $\blacksquare$ denotes continuous dense-phase, $\square$ denotes localized dense-phase, and $\triangle$ represents negligible cluster formation. The averages are taken for steady-state values, for $5$ independent simulations.}
\label{fig:2}
\end{figure}

We ensure that the size of the simulation box $L > 2R$ in the case of both static as well as motile walls. Keeping these constraints, we study different systems in which $N$ varies from $N = 725$ to $N=11345$ for static walls with $R$ ranging from $16$ to $64$. For motile walls, we choose $R=64$ and $N= 11345$. We observe that for static walls, the aggregate converge into a steady state at a relatively short time, $t < 20$ for most cases. Thus, we run our simulations until $t=100$ and analyze the steady-state behavior of the system. For a few cases at large $\delta$, where the system undergoes a transition from negligible clusters to LC, it takes relatively longer (upto $t=85$) to reach a steady state. To improve the averaging, each parameter values are examined from $5$ to $10$ independent simulations. For calculations involving cluster dynamics and pressure tensor, we run our simulation till $t = 1000$. 
We have employed Euler integration scheme, with time-step $10^{-5}$ for $Pe < 180$ and $5 \times 10^{-6}$ for $Pe \geq 180$. In the case of motile walls, the simulations are performed until $t = 4500$ to calculate the dynamical behavior of the system by averaging over $40$ independent runs. For the system with static circular wall, we numerically study eq.~\ref{eqn:1} for values of $Pe$ from $30$ to $240$, $R$ from $16$ to $64 $ and $\delta$ from $1.25$ to $2.05$ and determine the characteristics of dense-phase morphology. For motile walls, we simulate for various values of porosity (ranging from $1.25$ to $1.95$) by keeping the motility fixed at $Pe = 150$. 

% The simulations are conducted in a square box of length $L$ with periodic boundary conditions. For consistency the number density of the active bath  $\phi = N a /L^2$, where $a = \pi \sigma^2/4$, the effective area of a single particle~\cite{Redner2013,Fily2012} is maintained  $0.3$ for all simulations. Therefore for a box of length 85.3 and a ring with $R = 64 $ the total number of bath ABP's is 11345. At this number density, bulk phase separation far away from the wall does not take place substantially.\\
%We use $\sigma$ as the length scale for the system, and set it, $D$ and $\beta=1/k_{B}T$ to $1$. We consider $t= \sigma^{2}/D$ as the unit of time and $1\times10^{-5}t$ as the integration time step. We run our simulations for $100t$ to examine the aggregation upon the static ring and till $4500t$ to explore the properties of the dynamical ring. Because of the stochastic nature of simulations, the clustering characteristics and the motional trajectories of the moving ring significantly vary for different independent simulations for the same parameter values. Therefore we run simulations for $5-30$ independent trials before taking the averages.

\section{Aggregation of ABP's on static circular walls}

Our previous study on planar porous walls~\cite{Das2020} has shown that in the low porosity limit  ($\delta < 1$), the wall is practically impenetrable for the ABPs, causing a uniform dense-phase formation on the surface. On the contrary, when $\delta \gg 2$, the wall permits ABPs to penetrate through easily, causing negligible aggregation on the wall. However, at the intermediate porosity, $1.0<\delta\lesssim 2.0$ the steric hindrance due to the wall is significant, while the ABPs are still able to penetrate through, leading to interesting morphological properties of aggregates. Thus, we focus our analysis on this intermediate range of porosity. We observe similar behavior in the case of circular walls, wherein the dense-phase undergoes a transition from a connected dense-phase (CDP) (Fig~\ref{fig:1}(a)) (ESM-Movie1) to a localized cluster (LC) (Fig~\ref{fig:1}(b))(ESM Movie2) as we increase the wall porosity. This behavior has qualitative similarities to wetting-dewetting transition observed at equilibrium liquid-solid interfaces. We thus focus our analysis on characterizing such transitions quantitatively. 
\subsection{Effect of pore size and particle motility}
\begin{figure}
    \centering
    \includegraphics[width=0.72\columnwidth]{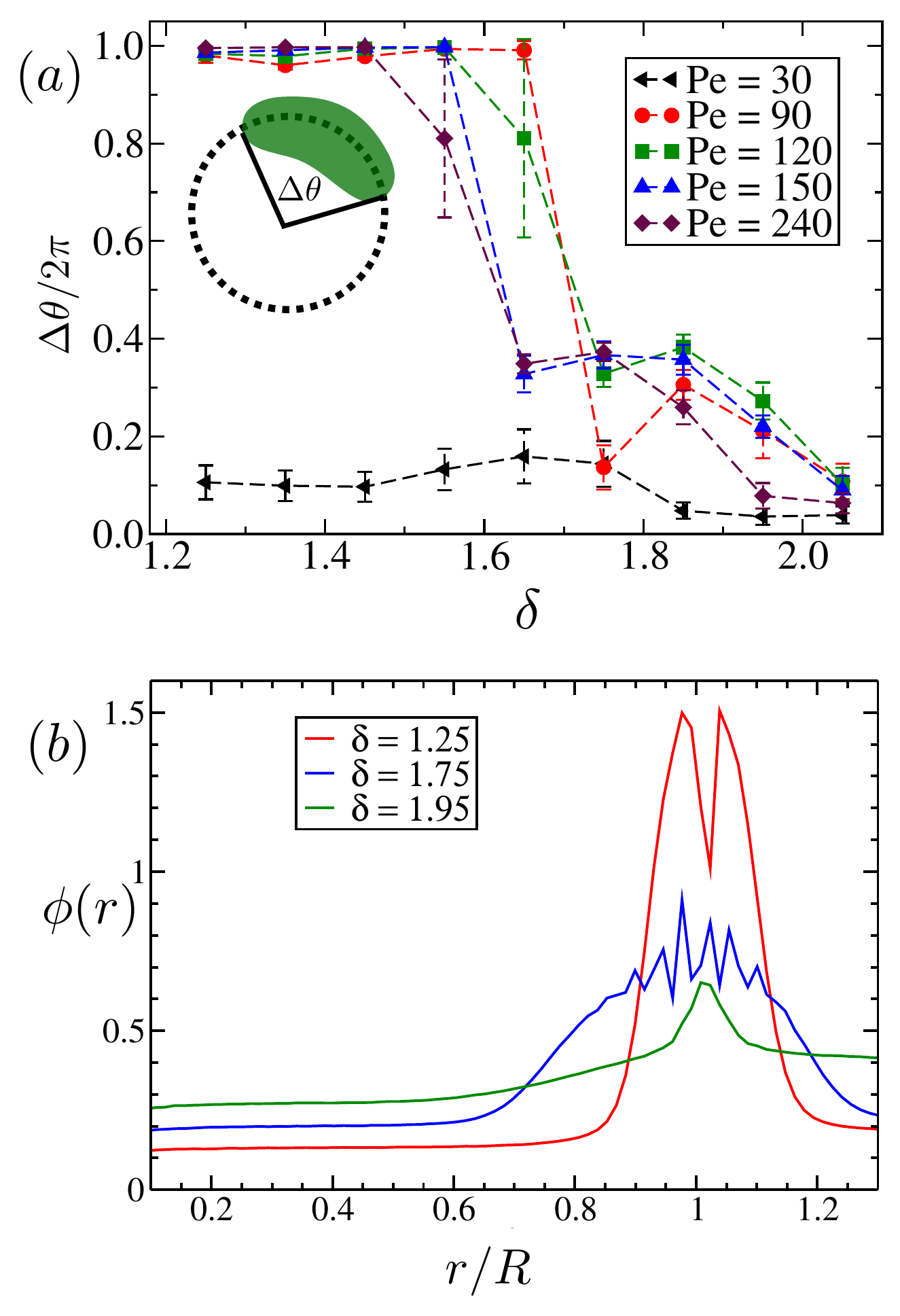}
    \caption{(a) The angular spread of the dense-phase, $\Delta \theta / 2\pi$ on the wall as a function of $\delta$ for different values of $Pe$. $\Delta \theta /2 \pi \simeq 1$ indicates the formation of CDP. The schematic represents the definition of $\Delta \theta$. The error bars are standard deviation, which have larger values near CDP-LC transition points.} (b) Local ABP density as a function of the radial distance, $\phi(r)$, from the center of the circle, shown for three different aggregation behavior corresponding to different wall porosity at $Pe = 150$. The radius of the ring, $R = 64$ for both the figures. The quantities are averaged for steady-state configurations for $5$ independent simulations.
    \label{fig:3}
\end{figure}

First, we systematically study the dependence of dense-phase morphology on particle motility and the wall porosity. For this purpose, we consider a ring of fixed radius, $R=64$ and analyze the ABP aggregation by changing $\delta$ from $1.25$ to $1.95$ and $Pe$ from $30$ to $240$. We observe the formation of dense-phase clusters for a range of values of $Pe$ and $\delta$.  To analyze this dense-phase quantitatively, we identify the particles which adhere to the wall and form clusters. We assume that an ABP has adhered to the wall if its separation from the wall particle is less than the interaction cut-off, $r_{\text{cut}} = 2^{1/6}\sigma$, which is also the definition for particles forming clusters. This estimation also allows us to calculate the cluster boundary and cluster thickness on the wall. This way, we compute the fraction of total ABPs, which are part of the dense-phase cluster on the wall, called cluster fraction $N_c$. In our simulations, $N_c$ reaches a steady state before $t<20$ for most values of ($Pe$,$\delta$) and $t< 85$ for a few cases at $\delta \geq 1.75$, where the transition from negligible clustering to LC takes place. We also repeat the simulation for five independent initial conditions for better averaging. The average cluster fraction, $\langle N_c \rangle$ acts as a good indicator for cluster formation. Here, $\langle. \rangle$ indicates the averaging over all the steady-state configurations in all the five independent simulations.  For $Pe < 30$, $\langle N_c \rangle \simeq 0$ for all values of $\delta$. However, for larger $Pe$, $\langle N_c \rangle$ reaches a steady nonzero value for a range of $\delta$. In Fig~\ref{fig:2} we show the $\langle N_c \rangle$ averaged over the steady-state values, as a function of $\delta$ for different $Pe$. We find that for $\delta \lesssim 1.5$, $\langle N_c \rangle$ saturates at a large value ($0.55$-$0.7$), with more than half of the ABPs becoming a part of the dense-phase. A qualitative examination reveals that the aggregate on the wall forms a CDP at this range of $\delta$, where $\langle N_c \rangle$ saturates at a large value. However, for $\delta \gtrsim 1.55$, $\langle N_c \rangle$ abruptly decreases to a lower value. At this range of $\delta$, instead of decreasing monotonously to zero, $\langle N_c \rangle$ shows a second plateau, where we qualitatively observe the formation of localized clusters. When $\delta > 1.9$, $N_c$ decreases rapidly again, indicating negligible dense-phase formation on the wall with $\langle N_c \rangle \simeq 0$. 

\begin{figure}
    \centering
    \includegraphics[width=0.9\columnwidth]{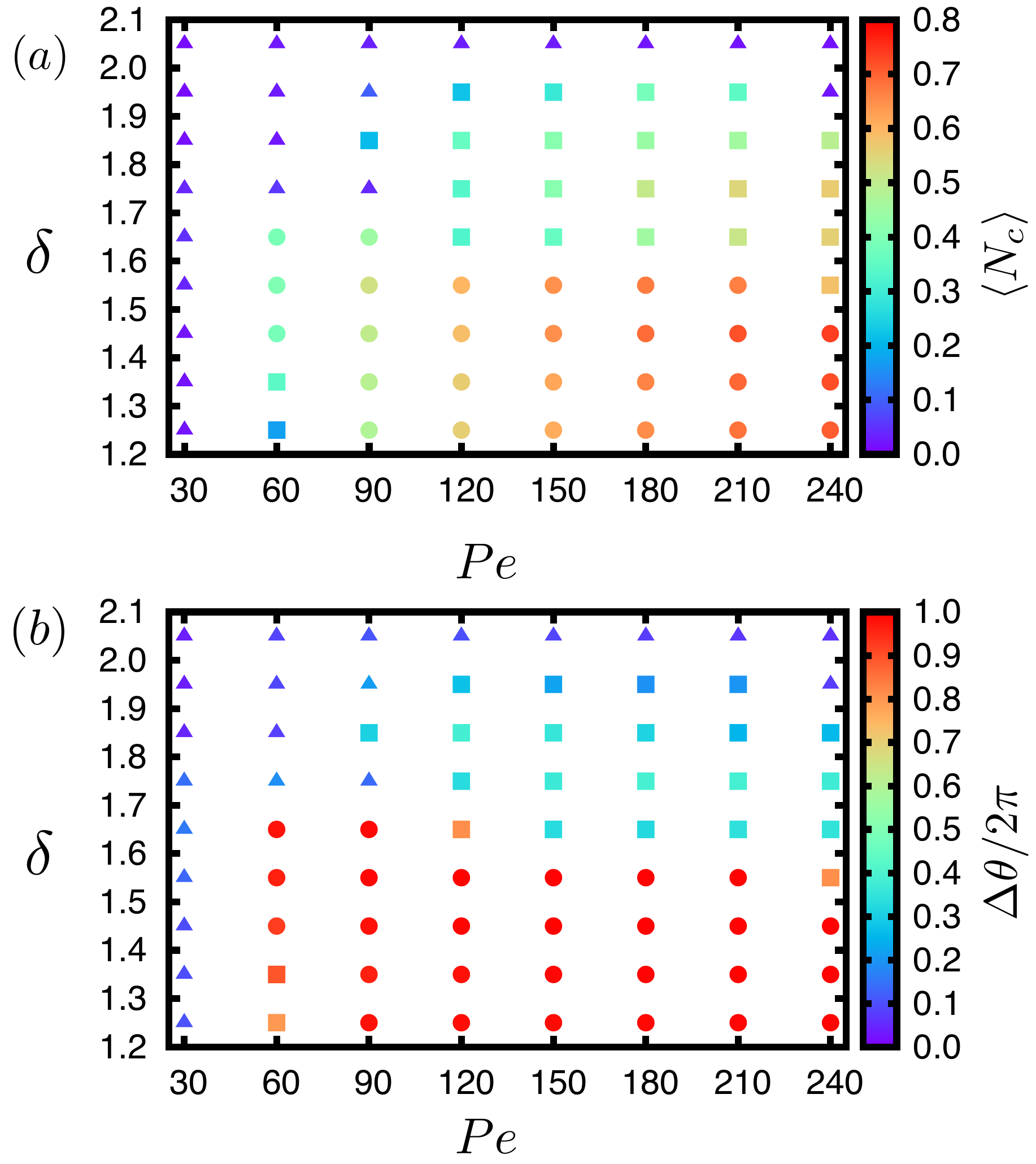}
    \caption{State diagram as a function of porosity $\delta$, and activity $Pe$. The color coding in panel (a) indicates the cluster fraction $\langle N_c \rangle$, while in panel (b) depicts the fraction of angular spread of the largest cluster, $\Delta \theta/2\pi$. The diagrams mark CDP ($\bullet$), LC ($\blacksquare$) and negligible clusters ($\blacktriangle$). Each points are averaged over steady state configurations, for $5$ independent simulations.
    }
    \label{fig:5}
\end{figure}
%To closely analyse the different cluster morphologies we plot the fraction of particles interacting with the wall and the fraction of particles in the largest cluster in Fig\ref{fig:2}(c). We plot for a high value of $Pe = 210$ where we distinctly observe transitions between the three different types of aggregation. We see for continuous clusters at $\delta < 1.55$ the fraction of particles interacting with the wall and the fractional size of the largest cluster is around $0.7$. For $1.55 \lesssim \delta \lesssim 1.85$ the number of particles interacting with the wall decreases, due to the increased porosity and therefore in this regime we see discrete clusters dominated by a single large cluster on the ring walls. Further increase in porosity at $\delta > 1.85$ shows the size of the largest cluster to decay to below the number of particles interacting with the wall. This is in keeping with the observations of negligible clustering and condensation upon the ring wall for very high porosity.\\
To draw a quantitative distinction between different dense-phase morphology, we calculate the normalized angular spread, $\Delta \theta /2\pi$ (Fig~\ref{fig:2}(a)) of the dense-phase on the circular wall. To calculate this quantity, we identify the largest cluster using the cluster definition defined previously and locate the edges of the cluster on the wall where the cluster thickness drops to zero. Such edges do not exist for CDP, hence $\Delta \theta = 2 \pi$. In Fig~\ref{fig:3}(a), we plot $\Delta \theta /2\pi$ as a function of $\delta$ for different $Pe$. When $\delta < 1.6$, the dense-phase form a CDP on the circular wall, where $\Delta \theta / 2\pi \simeq 1$ for large enough ABP motility ($Pe >30$). Around $\delta \simeq 1.6$, we observe a transition from CDP to LC where $\Delta \theta/ 2 \pi$ decreases abruptly from $1$ to a lower value (Fig.~\ref{fig:3}(a)). In this regime, the angular spread is $0.2$ to $0.4$. This transition is consistent with the quantitative estimation of $N_c$ in Fig~\ref{fig:2}. Further increasing the pore size $\delta > 1.85$, we find that the previously observed single large cluster disappears with ${ \Delta }\theta/2\pi \lesssim 0.2$ and the ring circumference is covered with several small clusters of rapidly fluctuating sizes. We also note, as previously mentioned, that for a very low value of $Pe = 30$, the system does not show significant phase separation as $N_c < 0.1$ for all values of $\delta$. However, at this parameter value, we observe a large number of local, transient clusters forming on the wall. These fluctuating clusters manifest as a nonzero $\Delta \theta$ and a weak nonmonotonicity with the change in $\delta$.

We also calculate the local ABP density as a function of the radial distance $r$ from the center of the circle, $\phi (r) = N(r) /\{4\pi\big((r + \Delta r)^{2} - r^{2}\big)\}$, with $N(r)$ being the number of particles in the shell between $r$ and $r +\Delta r$. In Fig~\ref{fig:3}(b), we show $\phi (r)$ for three values of $\delta$, which show distinct aggregation properties at $Pe=150$. For all these values of $\delta$, we observe $\phi(r)$ reaches a maximum as  $r\rightarrow R$. The peak value of $\phi$ is highest for the smallest wall porosity, where the aggregate form a CDP. In this regime, there is a dip in $\phi(r)$ at $r=R$, since a less porous wall does not permit any ABPs along the circle. For higher porosity ($\delta = 1.75$), $\phi(r)$ shows a  relatively broader distribution, while its peak becomes significantly smaller. This can be understood as for the LC, the radial width of the aggregate is higher compared to the clusters in the CDP phase, as evident from the comparison between Figs~\ref{fig:1}(a) and (b).
Also, we note that more ABPs are part of the dense-phase in CDP form, in comparison to LC, as evident from Fig.~\ref{fig:2}. Therefore, the density away from the wall is also relatively higher in the case of LC, as it reflects a larger number of ABPs in the vapour phase. Beyond $\delta = 1.95$, the local density is nearly constant with a small peak at $r = R$. As noted from Fig~\ref{fig:3}(b) with only a small number of particles in clusters, the localized distribution of particles $\phi(r)$ is more evenly distributed.
In Figs.~\ref{fig:5}(a) and \ref{fig:5}(b), we summarize different aggregate morphology observed in the parameter space defined by $\delta$ and $Pe$. In Fig.~\ref{fig:5}(a), we quantify the simulations in the parameter space with $\langle N_c \rangle$ to distinguish the regions where significant wall aggregation is observed. For $Pe \lesssim 30$ and $\delta \gtrsim 2.0$, we do not observe significant aggregation where $\langle N_c \rangle \leq 0.1$.  When $\delta \leq 1.95$, we observe a range of large $Pe$ where the aggregation takes place. Further, to distinguish different morphological states within the aggregate quantitatively, we use the angular spread $\Delta \theta / 2\pi$ in Fig~\ref{fig:5}(b). If $\Delta \theta /2\pi < 0.9$, we classify the cluster to be CDP. With this measure, we observe the formation of CDP at $\delta \leq 1.55$ for a range of $Pe$. Around this $\delta$ (near the transition regime) there is a tendency to form LC at higher $Pe$. However, for $\delta \lesssim 1.45$ we observe only CDP at large $Pe$. The aggregates form LC in regions between $1.55 \leq \delta <2.0$ at large $Pe$, as $\Delta \theta /2\pi$ displays a sharp decrease at these values. This region is clearly demarcated in the state diagram (Fig~\ref{fig:5}(b)) with a distinctly smaller value of $\Delta \theta / 2\pi$.  
\begin{figure}
    \centering
    \includegraphics[width=0.95\columnwidth]{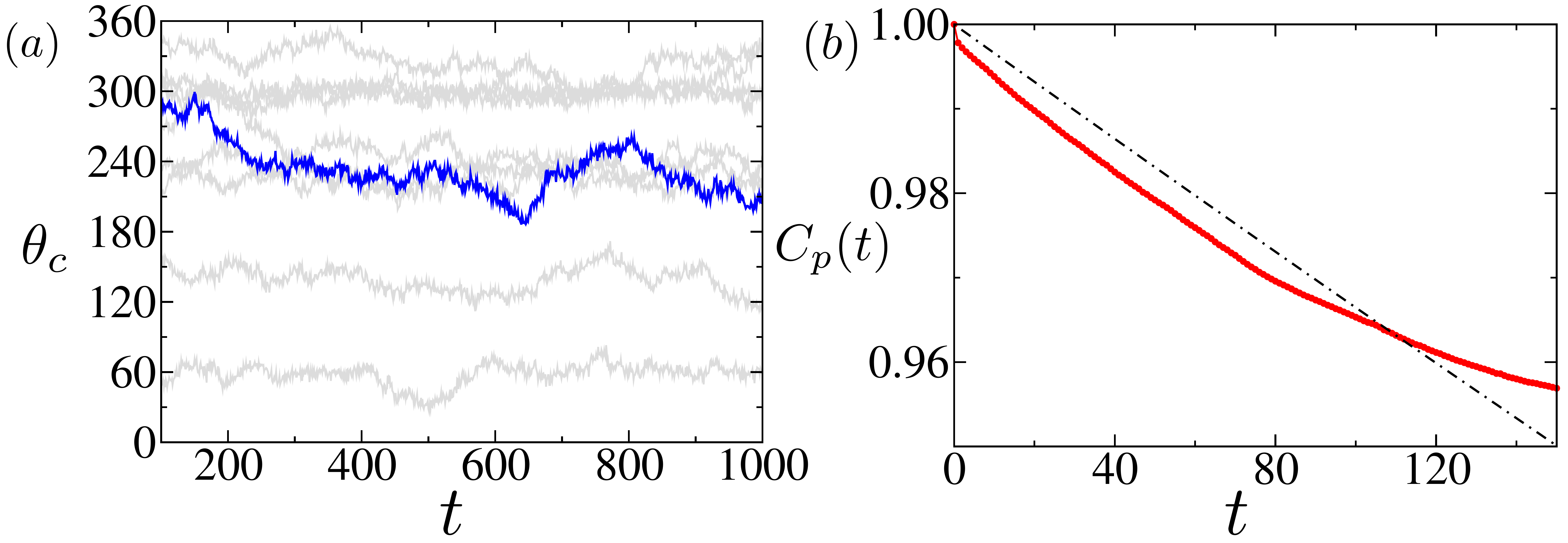}
    \caption{ (a) Variation of mean angular position $\theta_c$ with time. The gray lines are for the $10$ independent runs for $\delta = 1.75$ and $Pe = 150$, while the blue line indicates a typical time evolution behavior of $\theta_c$. (b) Auto-correlation, $C_p$ of the polarity vector ($ {\bf p} = \{ \cos\theta_c, \sin\theta_c \} $) of $\theta_c$ of the localised cluster for $\delta = 1.75$ and $Pe = 150$, averaged over $10$ independent runs (red filled circle). The black dotted line indicates the fit with a function $f(t) =  \exp (-t/\tau_0)$. We obtain the relaxation times, $\tau_0 = 2930$ is much larger compared to the observation time (t=1000). 
}
    \label{fig:4}
\end{figure}
\subsection{Time-evolution of the angular location of LC} 
Since the dense region is a nonequilibrium state, particles continuously join and escape at its boundary. Also, the particles inside the dense region show interesting dynamical behavior~\cite{Redner2013}. Due to these effects, we expect a phase-separated dense region to be nonstationary at large timescales, even though the system is in a steady-state in terms of $N_c$. However, in this particular system, the aggregates are confined only on the wall, and its overall movement, if at all possible, is constrained. Due to this property, the cluster translation is not observable for CDP on static walls. But in the case of LC, it may still be possible to observe a long-time translation of the aggregates along the wall. To study this dynamics, we calculated the angular displacement of the mean angular position $\theta_c(t)$, indicating the midpoint of the aggregate. In Fig~\ref{fig:4}(a), we plot the time-evolution of $\theta_c$. Although, $\theta_c(t)$ displays significant short-time fluctuations, it does not show a systematic drift over long time. Thus, the cluster dynamics is similar to fluctuations about a mean value. We further analyze this dynamical behavior by calculating the auto-correlation of the radial vector ${\bf p} = \{ \cos\theta_c, \sin\theta_c \}$. In Fig~\ref{fig:4}(b), we plot $C_p(t) = \langle {\bf p}(t_0).{\bf p}(t_0+t) \rangle_{t_0}$, averaged over ten independent simulations. The correlation function fits poorly to a single exponential decay; however, it provides a decay rate $\tau_0 \simeq 2930$, which is much larger than the observation time. 
The time evolution of $\theta_c$ suggests an absence of a large angular translation of LC on the circular wall, at least within the observed simulation time. However, the dynamics of the cluster is entirely reflected in the small timescale fluctuations in its angular position. The absence of large time displacement of LC indicates the importance of initial clustering at the wall due to the clogging of particles that move across the wall from both sides. Once the small cluster grows into a phase-separated region, the reduction of overall particle density outside the dense region prevents the formation of new clogged locations.  We also note that the single exponential decay may not describe the dynamics of cluster on the wall, which may involve multiple timescales. However, $\tau_0$ provides a rough estimate of the relevant timescale of the cluster dynamics. We also note that, a more detailed analysis of this property requires much longer simulations. 
\subsection{Stress distribution}
The change in morphology of ABP dense-phase around $\delta \simeq 1.6$ is similar to the CDP-droplet transition, which we have reported in a different system with planar walls~\cite{Das2020}. There, the transitions are rationalized by calculating the stress distribution across the wall and the interfaces, from the local pressure tensor components. In this particular active system, the pressure tensor has three contributions. The 'swim pressure' originating from self propulsion of the ABPs is given by, 
\begin{equation}
  p_{\alpha \beta}^{(s)}(h) = \dfrac{1}{2A_h}\langle \sum_{i \in A_h} j_{i\alpha}v_{i\beta}\rangle,
  \label{eq:swim}
\end{equation}

where $\langle.\rangle$ indicates time-average within the area element $A_h$ located at a distance $h$ from the wall and $\alpha,\beta$ indicate Cartesian components of the tensor. The active impulse $\bm{ j}_i$ is defined as $v_0{\hat{\bf{e}}}_{i}/\beta D D_r$~\cite{Fily_2017, das2019local}. Another major contribution comes from the inter-particle interaction $F_{ij}$, called the interaction pressure,
\begin{equation}
   p_{\alpha \beta}^{(I)}(h) = \dfrac{1}{2A_h} \langle \sum_{(i/j) \in A_h} F_{ij\alpha}r_{ij\beta}\rangle,
   \label{eq:inter}
\end{equation}
Similarly, the particle-wall interaction $F_{iw}$ contributes via the wall pressure,
 \begin{equation}
    p_{\alpha \beta}^{(w)}(h) = \dfrac{1}{A_h} \langle \sum_i F_{iw\alpha}r_{iw\beta}\rangle,
    \label{eq:wall}
 \end{equation}
where $r_{iw}$ is the distance between the particle $i$ and the wall. At the interface, the diagonal terms of the tensors can be defined as $p_N$ and $p_T$, namely the normal and tangential components to the interface. If all the interfaces are aligned parallel in a simple planar geometry, then the total surface tension $\gamma_{tot}$ can be calculated from the integral across the interfaces~\cite{Kirkwood1949,nijmeijer1990wetting},
\begin{equation}
  \gamma_{tot} = {1\over 2}\int_{-\infty}^{\infty} \{(p^{(I)}_{N}-p^{(I)}_{T})+(p^{(w)}_{N}-p^{(w)}_{T})+(p^{(s)}_N-p^{(s)}_T)\}  dh.
    \label{eq:5_n}   
\end{equation}
%In the case of planar walls, we have analyzed and separately compared the stress tensor contributions as a function of the distance $y$~\cite{Das2020}. 
When the aggregates form CDP on circular walls, one can define a local normal and tangent for both wall-liquid and liquid-vapour interfaces. Thus, the integration in Eq.~\ref{eq:5_n} needs to be performed radially, where $p_N$ and $p_T$ becomes local normal and tangent pressure components of the particular location on the circle. However, for the LC state, the local tangent to both the interfaces are not parallel. Moreover, unlike the planar case, we do not observe an 'unstable-CDP' where the CDP cluster destabilize after a finite time. Therefore, making a direct comparison of stress distributions between parameter regions that prefer different morphological states becomes difficult in the case of circular walls. 

\begin{figure}
    \centering
    \includegraphics[width=0.72\columnwidth]{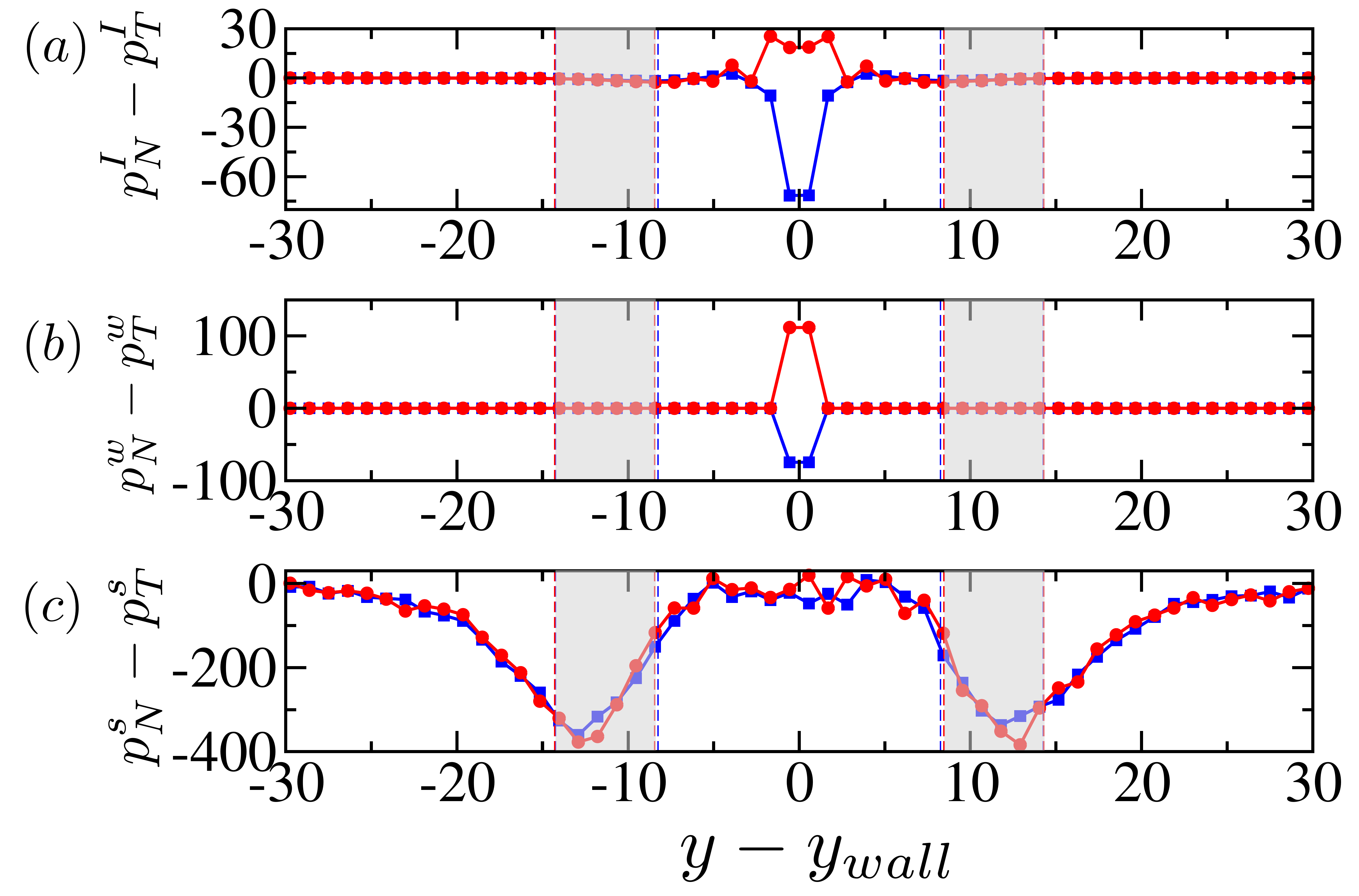}
    \caption{ Difference in normal and tangential components of (a) interaction, (b) wall, and (c) swim pressure as a function of vertical distance from wall, calculated for a system with planar walls with $N = 9900$, and $Pe = 180$. The blue curve is for $\delta = 1.25$ (CDP state) and the red curve is for$\delta = 1.85$ (unstable CDP), for $Pe = 180$. {The simulations are run  till $t = 1000$ for 10 independent runs. We took only CDP morphology at $\delta =1.85$ to calculate pressure tensor. 
    For $\delta = 1.25$, the averaging is performed over steady state values for 10 independent simulation. For $\delta = 1.85$ the averages were taken only for those configurations showing CDP (six out of ten runs). The shaded region depicts the liquid-vapour interface.}
    }
    \label{fig:6}
\end{figure}

 Here we assume that the underlying reason for the different morphological states is the same for both planar and circular walls. Using the simplicity of a system with planar walls, we demonstrate that the wall-liquid surface tension is different for parameters which prefer different morphological states.  
 In the planar geometry, the wall is aligned parallel to $x$ direction, thus $p_N$ and $p_T$ become $p_{yy}$ and $p_{xx}$ respectively. Similar to the circular walls, we observe the formation of CDP for $\delta <1.6$ and localized clusters for larger $\delta$ and at large $Pe$~\cite{Das2020}. However, for some parameters ($\delta = 1.85$ and $Pe=180$, for example) we also observe formation of `unstable' CDP which switches its morphological state within a finite time, while a fraction of runs shows relatively stable CDP at the same parameter values. Thus, we calculate the components $p_N$ and $p_T$ of all three pressure contributions separately using Eqns. ~\ref{eq:swim},\ref{eq:inter} and \ref{eq:wall} at $\delta = 1.25$, where the aggregates clearly prefer CDP, and $\delta =1.85$ at the same $Pe$ ($Pe = 180$). In Fig.~\ref{fig:6}, we compare the difference ($p_N - p_T$) for these two parameter values, as a function of the distance from the wall. We find a significant difference in interaction and wall pressure contributions, while the difference arising from the swim contribution is negligible. Also, $(p_N^{(I/w)}- p_T^{(I/w)}) <0$ at $\delta =1.25$ (CDP state) while it is positive at $\delta = 1.85$.  Since the difference $(p_N^{(s)}-p_T^{(s)})$ is not significant near the liquid-vapour interface, one can calculate the wall-liquid interfacial tension,
 \begin{equation}
    \gamma_{wl} = {1\over 2}\int_{-L/2}^{L/2} \{(p^{(I)}_{N}-p^{(I)}_{T})+(p^{(w)}_{N}-p^{(w)}_{T})\}  dy,
    \label{eq:wl}
\end{equation}
which provides a negative value for $\gamma_{wl}$ at $\delta = 1.25$ (CDP state, $\gamma_{wl} \simeq -184.5$) and a positive value at $\delta =1.85$ ($\gamma_{wl} \simeq 165.4$). This analysis hints the possibility that the inter-particle interaction, controlled by the local density and particle orientation, plays a crucial role determining the morphology of the ABP aggregate.
 
\subsection{Effect of wall radius, R and system size, $L$.}
\begin{figure}
	\includegraphics[width=0.48\textwidth]{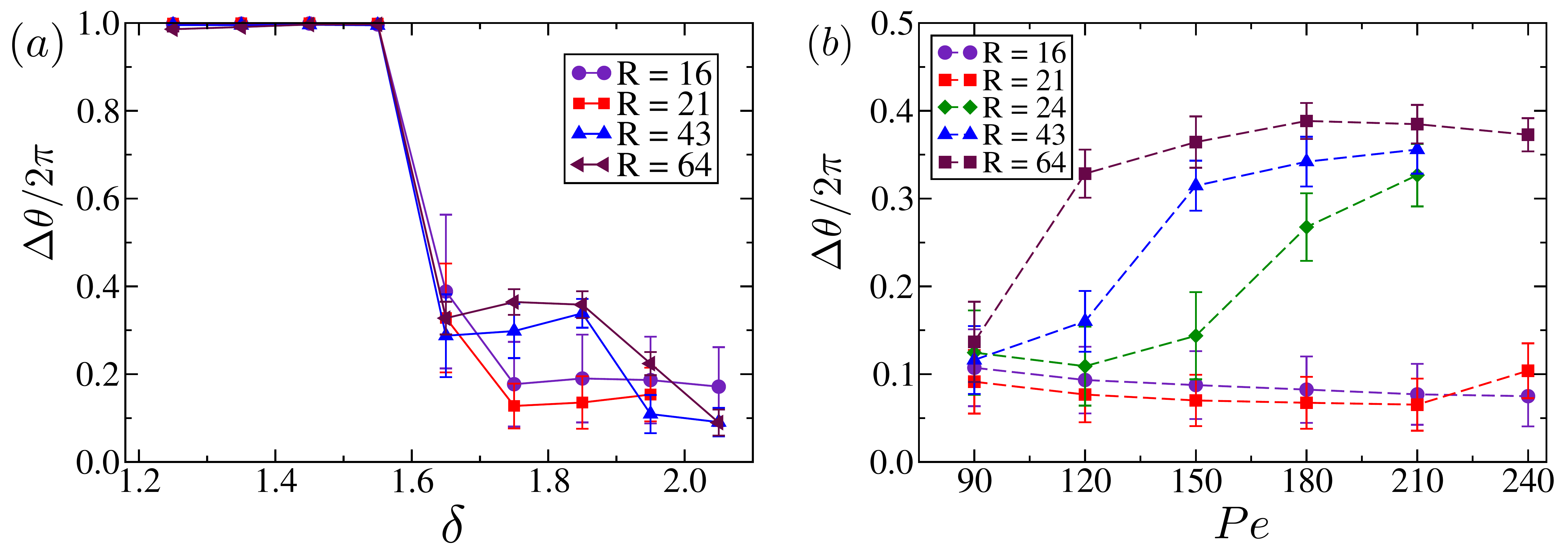}
\caption{The angular spread on the wall $\Delta \theta / 2\pi$ for various $R$ (a) as a function of $\delta$ at constant $Pe = 150$  (b) as a function of $Pe$ at a moderate porosity, $\delta = 1.75$. Below a critical radius we observe clustering to disappear for the localized clusters. The averages are performed over the steady-state values, for $5$ independent runs.} 
\label{fig:7}
\end{figure}
In the case of moderate porosity $1.6 < \delta < 2.0$, the aggregation at the wall takes place when particles from both sides of the wall meet at each pore, blocking each others passage. This mechanism requires an adequate number of particles both inside and outside the ring. Since the absolute number of particles inside the ring is proportional to the ring area, the aggregate properties can vary with $R$, even if all the other parameters are kept constant. Here, we examine the effect of the ring radius by analyzing the aggregation properties by varying $R$ from $16$ to $64$. In Fig~\ref{fig:7}(a) we show the angular spread $\Delta \theta$ of the dense-phase for different $R$, for a fixed particle motility ($Pe =150$), as a function of $\delta$. In the CDP regime ($\delta < 1.6$), $\Delta\theta$ is independent of $\delta$. Here, we do not observe any qualitative change in cluster morphology with a change in $R$.  However, in the LC regime ($1.65 < \delta <1.9$), we observe a reduction in $\Delta \theta$ when $R<22$, indicating a sensitivity to the value of $R$. 

To further analyze this, we fix the wall porosity at a moderate value $\delta = 1.75$ (LC regime) and vary the $Pe$. As shown in Fig~\ref{fig:7}(b), the behavior of $\Delta\theta$ as a function of $Pe$ qualitatively changes for $R<22$. In the cases of larger $R$ ($R>22$), $\Delta \theta$ increases monotonically with $Pe$, which is not the case for smaller rings (Fig.~\ref{fig:7}(b)). The low value of $\Delta \theta$ for small $R$ and at large $Pe$ indicates negligible dense-phase formation near the wall.
 This disappearance of dense-phase is expected, since $\phi$ for the entire system is constant and initially uniform in space, the number of particles inside the ring decreases as $R^{-2}$  whereas the maximum number of particles in contact with the wall decreases only as $R^{-1}$ as $R \rightarrow 0$. When $R=R*$, the interior of the circle has  a sufficient number of particles to block the pores, such that $n \phi\pi R*^{2} = 2\pi{R*}$, where $n$ is a constant indicating the fraction of particles attached to the wall in the steady state. For $R* = 22 $ we observe $n \approx 0.3$ which is comparable to the value of $N_c$, the fraction of total ABP part of the dense-phase. Increasing $Pe$ increases the penetration of the ABPs,  leading to less clogging at the wall.
\begin{figure}
\centering
\includegraphics[width=\columnwidth]{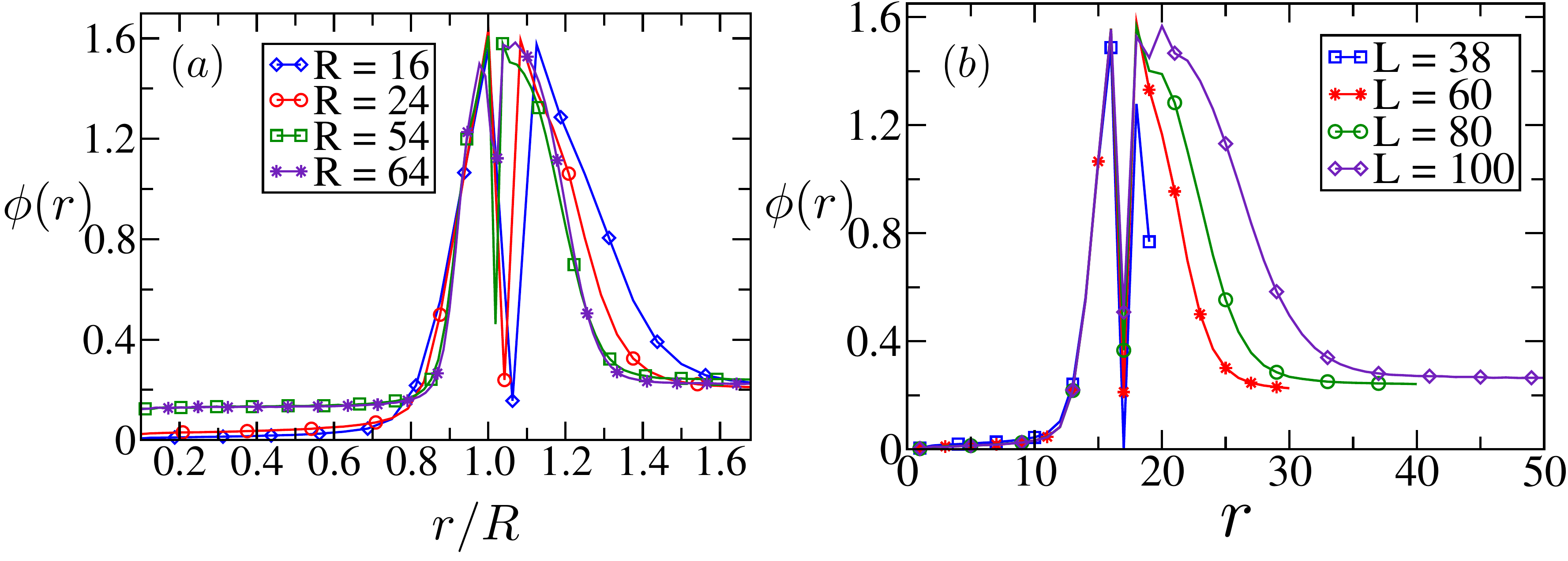}
\caption{a) The variation of radial density distribution $\phi(r)$ of ABPs as a function of the ring radius $r$ for $\delta =1.25$ and $Pe=150$. b) The effect of system size in radial density distribution, at $\delta=1.25$, $Pe=150$, $R=16$, plotted for different box-length $L$.}
\label{fig:8}
\end{figure}

However, at $\delta \lesssim 1.55$ (CDP formation), the ABP penetration through the wall is always very low, and therefore particles can accumulate on both sides independently. Thus, the clustering still occurs even at a small $R$ (analyzed here till $R = 16$). In Fig~\ref{fig:8}(a), we plot the radial density distribution $\phi(r)$ for $\delta =1.25$ and $Pe=150$ for different values of $R$. We maintain a fixed ratio of the ring radius to the box size with $R/L = 0.15$ in our simulations, to ensure a consistent ratio of ABPs inside and outside the ring. For large values of $R$, we observe that the density outside, far away from the wall approaches a constant value, while the $\phi(r)$ peaks as $r \rightarrow R$ in all cases. However, for $R\lesssim24$, $\phi(r \approx 0) $ is almost zero, since the majority of the particles inside the ring accumulate on the inner side of the ring.
%With a reduction of $R\lesssim24$ there is no clustering inside the ring apart from that on the inner wall.
%\paragraph*{\bf Finite Size effects.}
%\begin{figure}
%\centering
%\includegraphics[width=\columnwidth]{fig5.pdf}
%\caption{Increase in the box size, $L$, for a ring of radius $R=16$ leads to an increase in the particle density in outside the ring while the particle density inside the ring is independent of the box size.}
%\label{fig:9}
%\end{figure}
\begin{figure}
\includegraphics[width=\columnwidth]{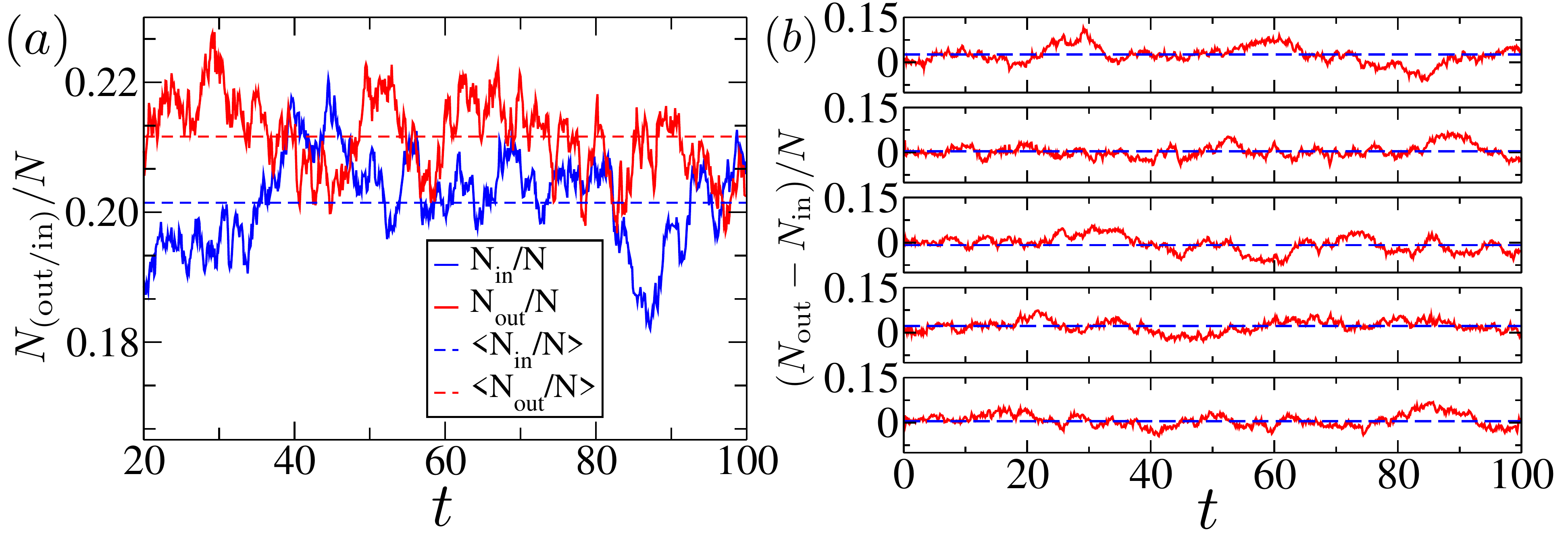}
\caption{(a) The fraction of ABPs outside (red) and inside (blue) the ring as a function of time (t). The dashed lines indicate the time-averaged value of these quantities. (b) The relative difference $(N_{\text{out}} - N_{\text{in}})/N$ in the number of particles inside and outside the ring for five individual runs. The number of particles on the outer wall exceeds the number of particles in the inner wall at most times. These simulations are done for $\delta = 1.75$ and $Pe =150$.}
\label{fig:9}
\end{figure}

We have also varied the simulation box size $L$ for a fixed value of $R=16$, $\delta=1.25$ and $\phi=0.3$, to analyze the effect of finite system-size. As seen in Fig \ref{fig:8}(b), the ABP density profile inside the ring remains constant for all values of $L$, as the total number of particles contained within the ring remains constant. However, an increase in the simulation box size leads to an increase in the aggregated particles in the ring exterior, leading to a concomitant increase in $\phi(r)$ at $r>R$.
\subsection{Asymmetry in aggregate deposition}

In our simulation, it is straightforward to examine the fraction of the total number of particles which aggregate on the inner and the outer surface of the circle. For a less porous ring, ABPs are confined to their respective regions inside and outside the boundary. Therefore, for a constant $\phi$, this distribution is a function of the ratio of the area of the two regions. This property is evident in Fig~\ref{fig:8}(b), where the value of $\phi(r)$ is higher outside
the wall at larger values of $L$. The situation for a porous wall is more interesting, in which case the particles cross from inside of the ring to outside and vice-versa. In Fig \ref{fig:9}, we plot separately the number of ABPs inside and outside the ring at $Pe = 150$ and $\delta = 1.75$, where the aggregate is in LC form. We observe an asymmetry in the number of ABP deposition since most of the time, a larger number of particles are deposited on the outer surface. This property is more clear in the time-averaged number of particles on the outer, as well as on the inner surface of the wall (Fig~\ref{fig:9})(a). In Fig~\ref{fig:9}(b), we plot the difference $(N_{\text{out}} - N_{\text{in}})/N$ for five independent simulations, which reaffirms the larger number of aggregates on the outer surface of the wall. This observation is significant as it implies that there is a net imbalance in the active force applied by the adhered particles on the wall, as every particle has the same propulsion speed.

%\paragraph*{\bf Deposition Asymmetry.}

%We examine the statistics of particle deposition on the inner and outer surfaces of the ring. For an impermeable ring, ABP's are confined to their respective regions inside and outside the boundary. Therefore, for a constant $\phi$, this distribution is a dependent on the ratio of the area of the two regions. More interesting is the situation for a porous wall permitting the transport of ABP's. We determine the particle deposition on the inner and outer ring surfaces. From Fig \ref{fig:9} we observe on average an asymmetry between the particle deposition on the inner and outer walls with often a higher deposition upon the exterior. As evident from Fig \ref{fig:9}(b) this excess deposition on the exterior surface was observed in four of our five independent simulations. Averaging over independent runs, we see in Fig \ref{fig:9}(a) the presence of a mean asymmetry. Since every particle has the same propulsion therefore the ring experiences a net active force. For a non static ring, the action of the net active force and its consequent displacement could be used for work extraction from the active bath.

\section{Dynamics of the moving ring}
\begin{figure}[h]
\centering
\includegraphics[width=0.45\textwidth]{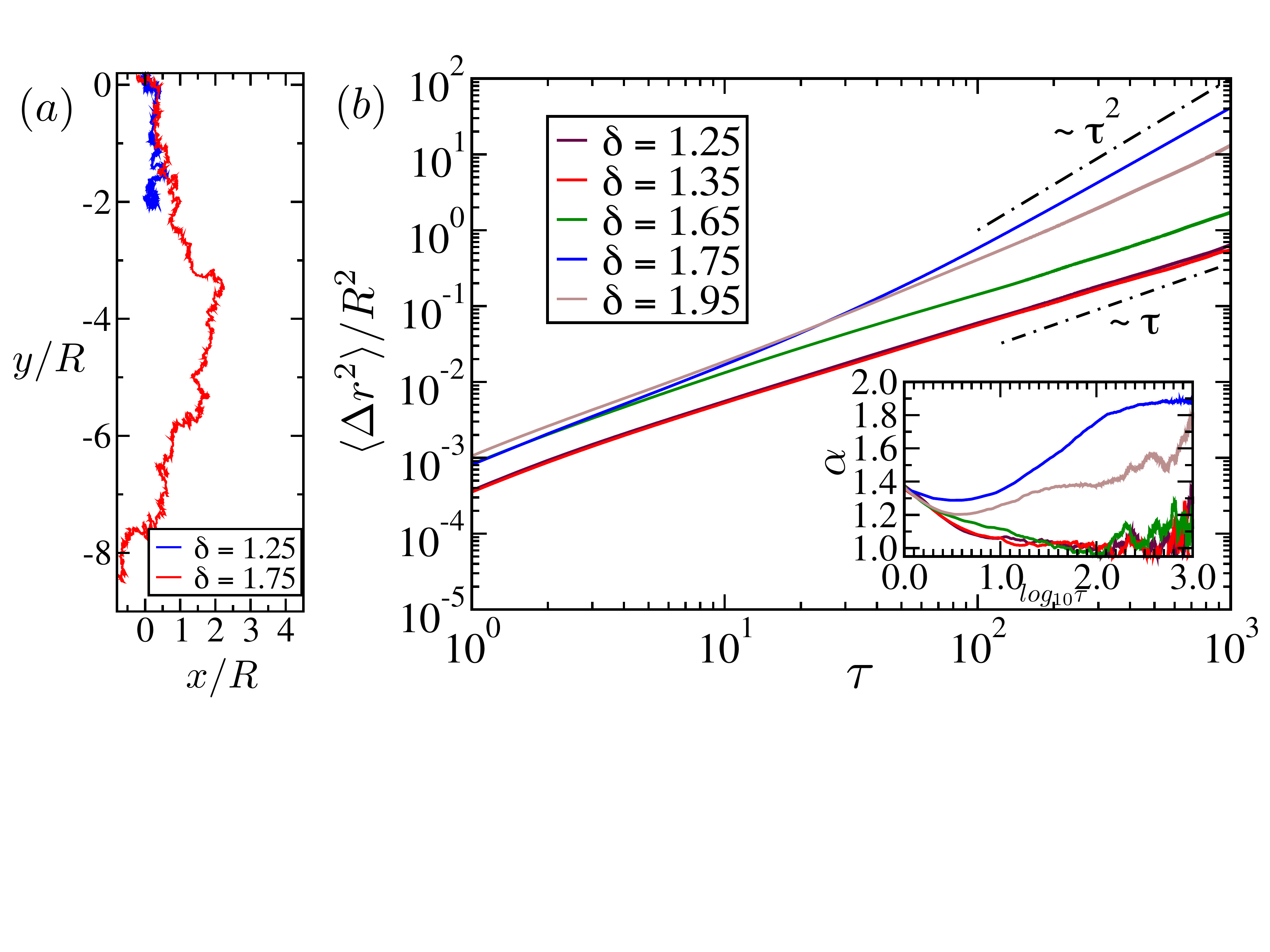}
\caption{(a) Typical trajectories of the ring for different ABP aggregation on the circumference for $Pe = 150$. The connected dense phase corresponding to $\delta = 1.25$ (blue), shows a more diffusive motion for the ring, whereas the localized cluster corresponding to $\delta = 1.75$  red), show directed motion at long timescale. (b) The mean-squared-displacement ($\langle \Delta r^2 (\tau) \rangle \sim \tau^\alpha$) of the center-of-mass of the ring for different wall porosity for $Pe = 150$.  (Inset) the exponent ${ \alpha} (\tau)$ from the mean-squared-displacement plot indicating diffusive behavior (${\alpha} \simeq 1$) for the connected dense phase and near ballistic behavior for the localized cluster (${\alpha} \simeq 2$). }
\label{fig:10}
\end{figure}

%\begin{figure}[h]
%\centering
%\includegraphics[width=0.45\textwidth]{fig8.pdf}
%\caption{Mean Squared Displacement of the ring for different ABP aggregation on the circumference for $Pe = 150$. The connected dense phase corresponding to $\delta = 1.25$ (colored in yellow), shows diffusive motion for the ring, whereas the droplets corresponding to $\delta = 1.75$ (colored in blue), show near ballistic behavior at long time scale. }
%\label{fig:8}
%\end{figure}
Our analysis shows that there is an imbalance in the number of particles that adhere to inside and outside of the circular ring. This imbalance can result in a nonvanishing net force when the dense-phase forms LC. When the circular ring is not restricted to move, the nonzero net force can cause its directed locomotion. Here we examine the possibility of using the aggregation anisotropy as a propulsion mechanism of a passive ring. The ring displaces as a rigid body due to the net force acted on its wall particles, using the method described in section II.  Keeping the radius constant at $R = 64$ and activity $Pe = 150$, we study the dynamics of the ring for $\delta$ ranging from $1.25$ to $1.95$. In the case of static walls, we observe CDP for $\delta  < 1.65$ and LC for higher values less than $1.95$. In Fig~\ref{fig:10}(a), we plot the trajectory of the center-of-mass of the ring up to time $t = 4500$ and compare it for $\delta = 1.25$ and $\delta =1.75$. Although the propulsion speed and the total number of ABPs are identical for both the cases, the ring with higher porosity covers a much larger distance at a given time. This difference in trajectories indicates that there exists a propulsion mechanism of porous rings when the dense phase forms LC. 

The difference in the dynamics of the ring for different $\delta$ is more evident when we compare mean-square-displacement ($\langle \Delta r^2 \rangle$) of the center-of-mass of the ring, where $\langle \Delta r^2 (\tau) \rangle = \langle ({\bf r}_{\text{cm}}(t+\tau) - {\bf r}_{\text{cm}}(t))^2\rangle$, $\langle . \rangle$ indicates both time and ensemble average ( Fig~\ref{fig:10}(b)). When the dense-phase form CDP ($\delta =1.25$, for example), the $\langle \Delta r^2 \rangle \sim t$, shows a diffusive dynamics, indicating the lack of long-term correlation in net forces acting on the ring. However, in the case of LC (for example, $\delta = 1.75$), the mean-square-displacement behavior is qualitatively different. In this case, we observe a diffusive behavior at low timescale, as in the case of CDP. However, at later times we observe a cross-over to a nearly ballistic dynamics ($\langle \Delta r^2 \rangle \sim t^2$), indicating a more correlated force acting on the ring. This behavior is more evident from the exponent ${\alpha} = \partial \ln(\langle \Delta r^2 \rangle)/\partial \tau $ (Fig.~\ref{fig:10}(b)). This type of cross-over from a diffusive to a near-ballistic dynamics is observed in different types of active systems, as well as passive tracer particles immersed in an active bath~\cite{Wu_PRL2000,chen2007fluctuations,leptos2009dynamics,wilson2011differential,mino2011enhanced}. Though the wall-particles are themselves not self-propelled, the coordinated behavior of the aggregates in case of LC leads to directed behavior. This behavior clearly indicates that it is possible to extract useful work from the bath of ABPs without inducing any shape anisotropy of the passive particle. We note that the self-propulsion is most effective when the dense-phase form a LC on the surface, at moderate values of $\delta$. 

\section{Summary}
We have studied the morphology of ABP aggregates on the presence of circular porous walls and studied its variation with wall porosity and particle mobility. We observe a qualitative change in aggregate morphology as the wall porosity is increased from a small to moderate value, from a continuously spread cluster to a localized cluster. At high porosity, the aggregation on the wall becomes negligible. We also show that aggregation behavior depends on the radius of the ring, especially at moderate porosity values when localized clusters are formed. At this range, the aggregation disappears when the ring radius is smaller than a critical value. We also observe an imbalance in the number of ABPs aggregated outside and inside of the ring in a localized cluster state. This imbalance causes a net nonvanishing force on the ring. We show that it is possible to make use of this net nonvanishing force on the ring to induce a directed motion and thereby extract useful work out of the system.  

Our numerical study is significant in understanding the collective spreading of self-propelled elements on substrates. The transition from a connected dense-phase to localized clusters is already reported in planar, porous walls in a previous study~\cite{Das2020}. The current study shows that such a transition also exists in the case of curved substrates.  We also show that the effect of the wall radius in dense-phase behavior is significant. However, to obtain a detailed understanding of morphological transition and to understand the origin of macroscopic length-scales in such systems, further analysis is required using combination of continuum and discrete models. Knowledge of such spreading behavior will be useful to design geometries that prevent the spreading of microbes, especially in the context of recent studies showing the significance of MIPS in microbial adhesion on substrates~\cite{Liu2019}. 
Interestingly, our study reveals the possibility of utilizing the aggregate geometry to extract useful work from a collection of self-propelled particles. So far, the focus of such studies is to determine optimal shapes to maximize useful work that can be extracted. We show the additional possibility of tuning the wall-particle interaction along with the wall geometry.

 \noindent{\bf Acknowledgements}
	 RC acknowledges the financial support from SERB, India via the grants SB/S2/RJN-051/2015 and SERB/F/10192/2017-2018. We acknowledge SpaceTime2 HPC facility at IIT Bombay.

%\begin{tabular}[h!]{ |p{3cm}||p{3cm}| }
% \hline
% \multicolumn{2}{|c|}{Table of Symbols} \\
% \hline
% $\sigma$   & Effective diameter \\
% $\delta$   & Ring Porosity \\
% $R$    &Radius of the ring \\
% $Pe$   &P\`eclet number \\
% $L$&   Box length  \\
% $d$& Arc length between two wall particles\\
% $D$& Translational diffusion coefficient\\
% $D_{R}$& Rotational diffusion coefficient\\
% $\Gamma(t)$& Gaussian Markovian random force\\
% $\mu$&Ring mobility\\
% \hline
%\end{tabular}

\bibliography{main}{}
\end{document}